\documentclass{aa}
\usepackage[varg]{txfonts}
\usepackage[utf8]{inputenc}
\usepackage{comment}
\usepackage{amsmath}
\usepackage{graphicx}
\usepackage{gensymb}
\usepackage[colorlinks=true,linkcolor=blue,citecolor=blue]{hyperref}
\usepackage{xspace}
\usepackage{geometry}  
\usepackage{lipsum}    

\usepackage{gensymb}
\DeclareRobustCommand{\ion}[2]{\textup{#1\,\textsc{\lowercase{#2}}}}
\newcommand{\jwst}{\emph{JWST}\xspace}
\newcommand{\ha}{H\(\alpha\)\xspace}
\newcommand{\Lya}{Ly\(\alpha\)\xspace}

\newcommand{\sirocco}{\textsc{Sirocco}\xspace}
\newcommand{\pyneb}{\textsc{PyNeb}\xspace}

\newcommand{\hei}{\ion{He}{i}\xspace}
\newcommand{\feii}{\ion{Fe}{ii}\xspace}
\newcommand{\oi}{\ion{O}{i}\xspace}

\usepackage{orcidlink}

\usepackage{natbib}
\bibpunct{(}{)}{;}{a}{}{,} 

\begin{document}

\title{Paschen jumps in little red dots: Evidence of nebular continua}
\authorrunning{Sneppen et al.}

\author{Albert Sneppen\inst{\ref{addr:DAWN},\ref{addr:jagtvej},\ref{addr:IoA},\ref{addr:kavli}}\orcidlink{0000-0002-5460-6126}, 
James H. Matthews\inst{\ref{addr:oxford}}\orcidlink{0000-0002-3493-7737}, 
Darach Watson\inst{\ref{addr:DAWN},\ref{addr:jagtvej}}\orcidlink{0000-0002-4465-8264}, 
Alex J. Cameron \inst{\ref{addr:DAWN},\ref{addr:jagtvej}}\orcidlink{0000-0002-0450-7306},
Stuart A. Sim\inst{\ref{addr:belfast}}\orcidlink{0000-0002-9774-1192}, \\
Joris Witstok\inst{\ref{addr:DAWN},\ref{addr:jagtvej}}\orcidlink{0000-0002-7595-121X}, 
Gabriel B. Brammer\inst{\ref{addr:DAWN},\ref{addr:jagtvej}}\orcidlink{0000-0003-2680-005X},
Kasper E. Heintz\inst{\ref{addr:DAWN},\ref{addr:jagtvej}}\orcidlink{0000-0002-9389-7413},
Georgios Nikopoulos\inst{\ref{addr:DAWN},\ref{addr:jagtvej}}\orcidlink{0009-0004-6791-9246}
}

\institute{
Cosmic Dawn Center (DAWN)\label{addr:DAWN}
\and
Niels Bohr Institute, University of Copenhagen, Jagtvej 128, DK-2200, Copenhagen N, Denmark \label{addr:jagtvej} 
\and
Institute of Astronomy, University of Cambridge, Madingley Road, Cambridge, CB3 0HA, UK \label{addr:IoA} 
\and 
Kavli Institute for Cosmology, University of Cambridge, Madingley Road, Cambridge, CB3 0HA, UK \label{addr:kavli}
\and
Astrophysics, Department of Physics, University of Oxford, 
Oxford OX1 3RH, UK\label{addr:oxford}
\and
School of Mathematics and Physics, Astrophysics Research Centre, Queen's University Belfast, Belfast, United Kingdom\label{addr:belfast}
}

\date{Received date /
Accepted date }

\abstract{
    Little red dots (LRDs) are broad-line sources at high redshift that were initially identified by their compact morphologies, red colours, and inflection near the Balmer wavelength. The origin of their optical-to-near-infrared continua is debated, with proposed explanations ranging from direct recombination emission to thermalised blackbodies from stellar-like atmospheres. We report evidence of Paschen jumps in a subset of LRDs with a high signal-to-noise ratio, consistent with free-bound recombination to hydrogen $n=3$. The Paschen and Brackett continuum shapes in the LRD sample are consistent with minimally reddened nebular emission from low-temperature gas with $T_e\lesssim10\,000$\,K, while the Paschen jump signatures limits scenarios in which the emission is thermalised. Further, the extreme H$\alpha$ equivalent widths and the tight observed correlation between H$\alpha$ and the continuum follow naturally if both originate in recombination emission. This provides an observational upper limit on the contribution of any direct active galactic nucleus accretion component and any stellar-atmosphere-like component, and also on the fraction of line emission that can be thermalised as it traverses the cocoon. Ultimately, ionised gas emission provide a self-consistent explanation of the continuum, line strengths, and line profiles without requiring multiple separately fitted components. \newline
    }
\keywords{}

\maketitle
\section{Introduction}
The population of compact, extremely red sources called `little red dots' (LRDs), which apparently are common at high redshift, has been one of the major discoveries of the James Webb Space Telescope (\jwst). They are characterised by broad hydrogen and helium permitted lines \citep{Matthee2024,Greene2024,Kocevski2024,Killi2024} as well as high surface brightness \citep{Kokorev2024,deGraaff2025}, initially suggesting virial motion around and accretion onto a supermassive black hole. However, their spectral signatures are more commonly associated with stars and transients. These include i) Balmer breaks and Balmer-line absorptions from large columns of hydrogen gas with significant $n=2$ populations \citep{Setton2024,Inayoshi2025,Barro2025}, ii) line broadening likely dominated by electron scattering \citep{Rusakov2025,Torralba2025,Sneppen2026}, and iii) red spectral continua with shapes reminiscent of blackbodies \citep{Kido2025,Liu2025,deGraaff2025b,Sun2026}.

These features collectively indicate that a dense stratified cocoon or envelope surrounds the supermassive black hole that contains predominantly ionised and predominantly neutral regions that reprocess the accretion power into red emission. Further, this cocoon might be able to suppress radio and X-ray emission, which are typically undetected \citep{Juodzbalis2024_rosetta,Maiolino_Chandra_2024,Sneppen2026c}, the high densities might naturally explain observed LRD line ratios deviating from case B recombination \citep{Nikopoulos2025,DEugenio2025b,Yan2025}, and the cocoon can reach temperatures $<3000\,$K permitting cold gas features \citep[e.g.,][]{Lin2025,Wang2026}. This motivates detailed radiative transfer modelling to capture this new physical regime and explain the puzzling nature of the red optical to near-infrared continuum \citep{Sneppen2026,Liu2026,Martins2026}. 

\begin{figure*}
\begin{center}
    \includegraphics[angle=0,width=0.95\textwidth]{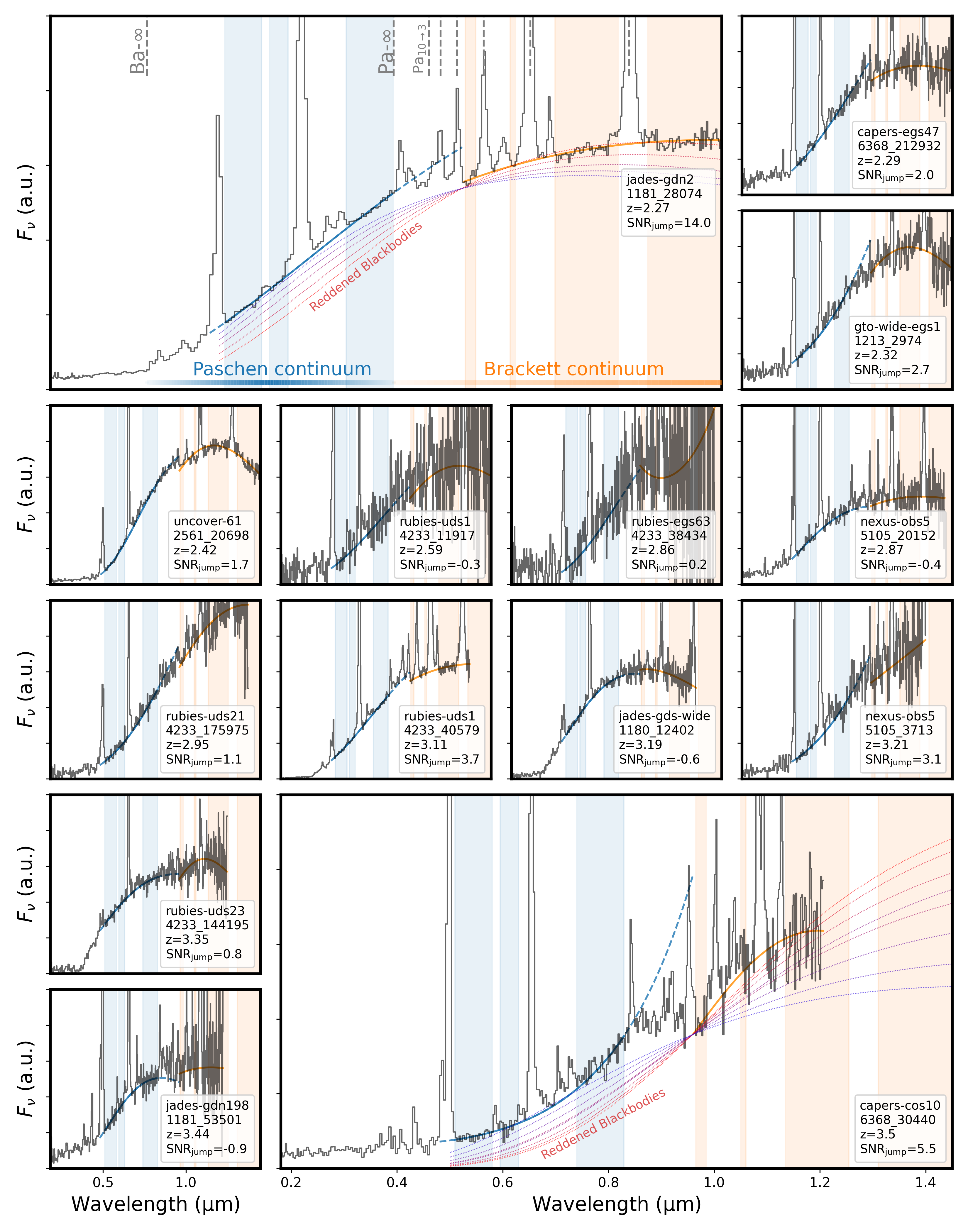}
\end{center}
\vspace{-0.6cm}
\caption{
\jwst NIRSpec/PRISM sample of LRDs with polynomial fits to the spectral continuum blueward (blue) and redward (orange) of the Pa-$\infty$ wavelength. The LRD data and \sirocco recombination models show a change in spectral slope at Ba-$\infty$ and Pa-$\infty$. The strong detected emission from high-order Paschen lines (e.g.\ Pa$\zeta$) implies blended recombination-line emission immediately blueward of Pa-$\infty$ (i.e.\ $n>10\rightarrow n=3$), so we fit the Brackett continuum at wavelengths with $\ge0.96\,\mathrm{\mu m}$. For each object, we provide in the legend the S/N difference between the Paschen extrapolated continuum and the Brackett continuum, where the former typically overshoots the latter (e.g.\ CAPERS-6368-30440). The reddened blackbody models have a smooth curvature and cannot match the Brackett and Paschen continua simultaneously (dashed red and blue lines).   }
\label{fig:paschen_sample}
\end{figure*}

Interpretations of this red continuum include accretion disc emission associated with active galactic nuclei (AGN) attenuated by large gas columns \citep{Inayoshi2025}, recombination emission of the gas cocoon \citep{Rusakov2025,Sneppen2026}, Hayashi-like gas envelopes \citep{Kido2025,Begelman2025} and thermalised blackbody-like emission from dense gas in the so-called `black hole star' model \citep{Naidu2025,deGraaff2025,deGraaff2025b,Sun2026,Liu2026}. While these proposed explanations draw upon the same idea of reprocessing by dense gas, they differ in whether the continuum is interpreted as stellar-photosphere–like emission versus partially processed recombination radiation. For instance, \cite{deGraaff2025b} provided an empirical study with a NIRSpec/\jwst compilation of 146 LRDs showing that their spectral continua can be reproduced with blackbodies with $T\sim2000{-}6000\,\mathrm{K}$ multiplied by a power law $\nu^{n}$ with index ${n\approx\pm3}$. These modified blackbodies are interpreted as coming from an optically thick envelope in approximate hydrostatic equilibrium. Extending this idea, \cite{Liu2026} provided a synthetic spectral library of modified blackbodies for LRDs computed using stellar atmosphere modelling codes for models with large column densities $N_{\rm H}\sim10^{27} {\rm cm^{-2}}$. By contrast, \cite{Sneppen2026} provided self-consistent 2D radiative transfer models in which the continuum is produced by recombination emission of ionised gas (i.e.\ a `nebular' continuum) rather than by a deeply thermalised atmosphere. Here, nebular is not meant to imply optically thin conditions. The \sirocco models are generically optically thick to electron scattering and their H$\alpha$ equivalent widths fall well below those of a pure Case~B. Synthetic spectra from these models yield a broadly similar continuum shape to a blackbody, but with one key observable distinction being a predicted change in spectral slope at the Paschen limit, Pa-$\infty$. If such spectral signatures can be detected (or excluded), it would allow us to probe the degree of thermalisation of the spectra. Nebular or diffuse continuum emission from broad-line region clouds can contribute to the optical continuum of standard AGN. In particular, Balmer-continuum emission is commonly associated with the small blue bump and the features of wavelength-dependent continuum reverberation lags \citep{Korista2001,Korista2019,Netzer2022}. Our point therefore is not that the underlying recombination physics is unique to LRDs. The question is instead whether LRDs occupy a regime in which the recombination continuum, including the Paschen continuum, remains visible and can dominate the total emergent spectrum. In nearby unobscured type 1 AGN, direct searches generally find smooth total-flux spectra across the Paschen limit, even when allowing for a substantial diffuse-continuum contribution \citep{Guo2022}.

While the Balmer break has revealed substantial surrounding column densities of hydrogen in $n=2$ \citep[e.g.][]{Inayoshi2025}, the Paschen continuum offers complementary information by probing the intrinsic emission of the ionised cocoon, the electron temperature through free-bound emission, and the degree of external reprocessing. We therefore explore the complementary lessons from Paschen jump signatures in LRDs. 
In Sect.~\ref{sec:obs} we summarise the \jwst NIRSpec/PRISM data, which include additional coverage up to 5.5 $\mu$m, which is useful for constraining the Brackett continuum at $z\leq3.5$. In Sect.~\ref{sec:model} we show how \sirocco modelling motivates the expected appearance of Paschen jump signatures in the regime of strong scattering and bright Paschen recombination-line emission. In Sect.~\ref{sec:results} we find and quantify these features of free-bound emission structures.
Additionally, we discuss in Sect.~\ref{sec:results} that pure blackbody thermalisation would require a distinct broad-line region outside the envelope photosphere. By contrast, the recombination framework in \citet{Sneppen2026} allows a single, continuous, and stratified medium to reproduce the key observables of LRDs.

\section{Methods}

\subsection{Observations}\label{sec:obs}
The data we used were processed via the DAWN JWST Archive\footnote{\href{https://dawn-cph.github.io/dja}{https://dawn-cph.github.io/dja}} (DJA) \textsc{v4.4} \citep{BrammerValentino2025DJA_NIRSpec}. The processing is described in \cite{DeGraaff2024_RUBIES,Heintz2025} with recent \textsc{v4} improvements detailed by \cite{Pollock2025,Valentino2025}. In particular, \textsc{v4} added wavelength coverage extending to $\lambda_{\rm obs}\sim\,$5.5$\,\mu$m, which is useful for constraining the continuum redward of the Paschen limit for the high-redshift LRD population. LRDs display strong emission features immediately redward of the Paschen limit, which necessitates wavelength coverage extending beyond $\lambda_{\rm rest}\gtrsim1.2\,\mu$m for strong constraints on the spectral shape. This limits the well-constrained Brackett continuum domain to the lower-redshift tail of the LRD population $z\lesssim3.5$, while the general population predominantly lies at $z\gtrsim4$ \citep{Kocevski2024}. Therefore, we analysed the $z\leq3.5$ subset of the NIRSpec/PRISM LRD population using the selection of \cite{deGraaff2025b}, who selected on i) the spectrum displaying an inflection near the Balmer limit and ii) being a compact source in imaging. This yielded 14 objects with a typical resolving power of $\mathcal{R}\sim100-200$ from $2.27\leq z\leq3.5$ (see Fig.~\ref{fig:paschen_sample}). In luminosity and Balmer-break strength, this low-redshift subset is representative of the broader LRD population, although it generally lacks the grating spectroscopy available for the sample compiled in our previous paper \citep{Sneppen2026}. We highlight the lowest-redshift object of the sample, the Rosetta Stone LRD (JADES~1181-28074), because it has been studied in detail \citep{Juodzbalis2024_rosetta,Brazzini2025,Brazzini2026} and provides a high signal-to-noise (S/N) test of the predicted Paschen jump structure.

\begin{figure}
\begin{center}
    \includegraphics[angle=0,width=0.45\textwidth]{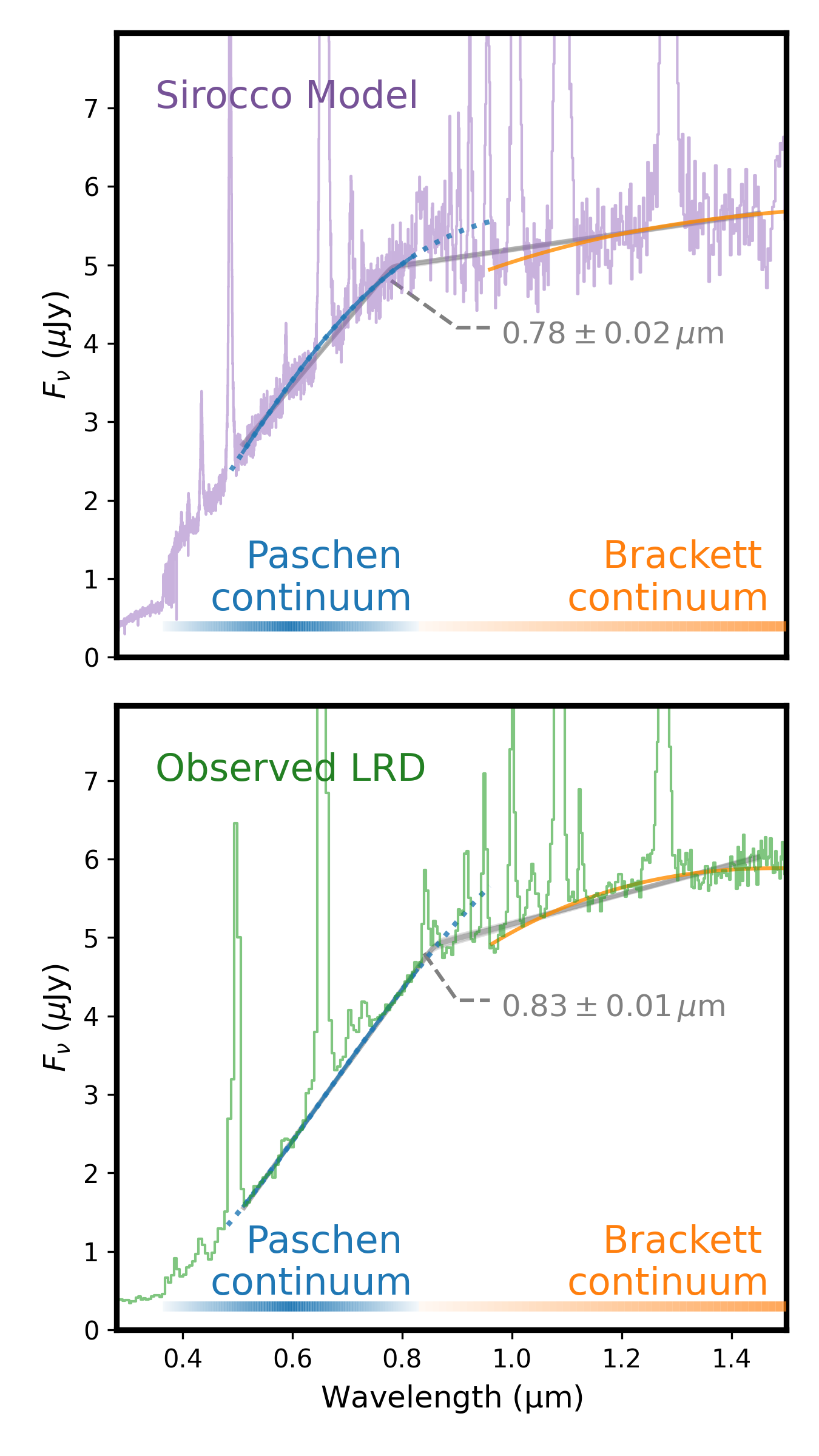}
\end{center}
\vspace{-0.6cm}
\caption{
Top panel: \sirocco LRD model highlighting the change in spectral slope expected around Pa-$\infty$. Bottom panel: Rosetta Stone LRD is shown for observational comparison. The polynomial fits to Paschen and Brackett continua are indicated in blue and orange, respectively, while a piecewise linear fit to both continua displays a change in spectral slope near Pa-$\infty$. The observed and \sirocco spectra show a continuum reminiscent of a blackbody although they are not thermalised, together with broad recombination lines from species with high ionisation potentials, such as \hei\,$\lambda 1083.3$\,nm.
}
\label{fig:sirocco_ex}
\end{figure}

\subsection{Sirocco modelling: Signatures of LRD Paschen jumps}\label{sec:model}
We used the Sobolev Monte Carlo code \sirocco \citep[{\it Simulating Ionisation and Radiation in Outflows Created by Compact Objects};][]{long2002,Matthews2025} for the radiative transfer modelling of LRDs, a problem that requires an accurate treatment of electron scattering, line transfer, and bound-free processes within optically thick multi-dimensional flows \citep{Sneppen2026}. For a given input outflow structure (mass density and velocity field), \sirocco computes the cocoon radiation field and ionisation structure, thereby determining the electron column density, the strength of the recombination lines and continua, the extent of partially ionised material, and the strengths of Balmer and Paschen absorption lines and breaks. \sirocco thus permits us to quantify and simultaneously reproduce the overall spectral shapes, line profiles, line and continuum luminosities, individual absorption features, ratios of the different lines, and correlations between these quantities \citep{Sneppen2026}.

\begin{figure}
\begin{center}
    \includegraphics[angle=0,width=0.5\textwidth]{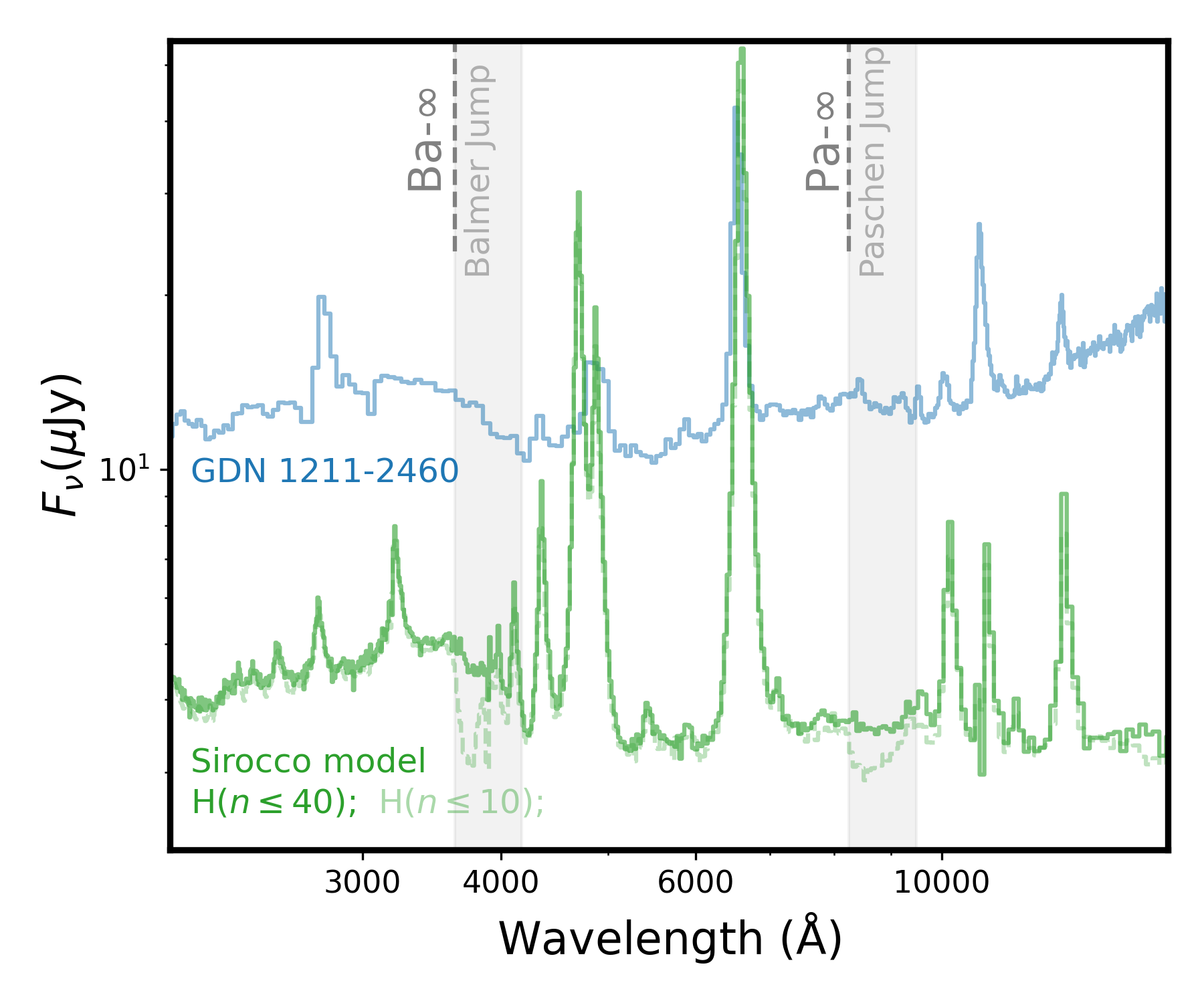}
\end{center}
\vspace{-0.6cm}
\caption{
$z=2.60$ broad-line compact little blue dot GDN 1211-2460 \citep{Maseda2024} displaying separated spectral slopes in the Balmer, Paschen, and Brackett continua. We show \sirocco modelling (green). The shading indicates that $n> 10$ energy-levels of hydrogen must be included for producing more accurate Balmer- and Paschen jump shapes. In the $n=40$ model and data, the Paschen jump is blended with closely spaced recombination line emission. 
}
\label{fig:LRD_example}
\end{figure}

The basic geometry we considered was that of a spherically symmetric cocoon with column densities of $N_H\sim10^{25}\,\mathrm{cm^{-2}}$, outflow velocities of $100\,\mathrm{km\,s^{-1}}$, and a central blackbody source. We focused on the broad-scale continua of these \sirocco model spectra at the NIRSpec/prism resolution, for which the detailed kinematic treatment required to model grating-resolved line structure in \citet{Sneppen2026} is less important. Although this earlier analysis found evidence of non-spherical kinematics from the line profiles, such effects are not central to the continuum-based discussion presented here, so we adopted the simpler geometry. The resulting simulation operated using the Sobolev approximation with line transfer treated in the hybrid macro-atom limit and computed the emerging spectrum after the ionisation and temperature structure had converged under the assumption of radiative near-equilibrium. The \sirocco results and modelling are discussed in more detail by \citet{Sneppen2026}, who already predicted the presence of Paschen jumps. We supplement the \sirocco H dataset and present two models with blackbody luminosities and temperatures of $3\times10^{44}\,\mathrm{erg\,s^{-1}}$ and $20\,000$ K, and $10^{45} \,\mathrm{erg\,s^{-1}}$ and $100\,000$ K, respectively, with similar cocoon column densities. The spectra from these models are shown in Fig.~\ref{fig:sirocco_ex} and Fig.~\ref{fig:LRD_example}, but they are not outliers in the simulation grid. Specifically, for a high-fidelity reproduction of the Paschen structure, we expanded the fiducial atomic dataset for H in \sirocco to include all levels up to $n=40$ using energy levels and A-values from NIST \citep{Kramida2023}, and we assumed the same photoionisation cross-sections for all levels with $n\geq20$. When applying dust extinction in this work, we used the SMC extinction curve \citep{Gordon2003}. 

A notable \sirocco prediction from \citet{Sneppen2026} is that recombination emission from ionised cocoons with columns of about $N_{\rm H}\sim10^{25}\,{\rm cm^{-2}}$, which is substantially below the much larger columns often invoked in fully thermalising atmosphere models, leaves a characteristic spectral imprint near 
$\lambda_{\rm rest}\simeq 820\,\mathrm{nm}$, the Paschen limit (Pa-$\infty$). 
No clear step-like jump in flux is expected from \sirocco models; instead, the key imprint is an acute change in spectral slope across Pa-$\infty$ (Fig.~\ref{fig:LRD_example}). 
The appearance of this free-bound signature in model spectra is subtle because of the strong, broad, and blended emission immediately redward of Pa-$\infty$ (see Fig.~\ref{fig:sirocco_ex}). In the synthetic \sirocco spectra, the blended emission component is dominated by high-order Paschen lines (as discussed in AGN by \citealt{Guo2022}), while potential contributions from \oi\ and \feii\ emission have also been noted \citep{Kokorev2025,Sun2026}. The LRD test we performed was whether the recombination continuum was sufficiently dominant for its bound-free spectral curvature to remain visible in the observed total continuum.
In practice, extrapolating the Paschen (blueward) continuum to the Brackett (redward) continuum yields a mismatch in flux level, with the blue-side extrapolation overshooting the red-side continuum. Fig.~\ref{fig:sirocco_ex} shows second-order polynomial fits to the Paschen and Brackett continua fitted in logarithmic space, where the models and observations both display a mismatch when extrapolated across the edge. We did not explore fitting with the \sirocco model spectra, as full Monte Carlo radiative transfer calculations are computationally expensive, with a \sirocco spectrum requiring about $10^2$ CPU hours. We instead focused on a prediction of \sirocco model spectra that we used to motivate this slope-change as our operational signature of Paschen free-bound emission and tested for it in the data.

\subsection{Fitting framework}
The majority of the 14 observed LRDs show Paschen-inferred continua whose extrapolation exceeds the Brackett continuum. To evaluate the significance of this excess and to provide clarity on the change in spectral slope, we explored various parametric models within a Bayesian framework. For each model, we sampled the posterior distribution of the parameters using the Markov chain Monte Carlo ensemble sampler \texttt{emcee} \citep{ForemanMackey2013_emcee}, adopting uniform priors over the parameter ranges described below. 

First, we provided an agnostic parametrisation of the continuum blue- and redward of Pa-$\infty$ with a second-order polynomial fitted to the logarithmic space of wavelength and flux, $\log_{10}(F_{\nu})=a\log_{10}(\lambda)^2+b\log_{10}(\lambda)+c$, which captures curvature without imposing a physical model. This allowed us freedom to reproduce smooth functional rollovers, or, on the other hand, to determine points of sudden spectral transitions. The wavelength regions we used for fitting the spectral continuum were chosen to avoid contamination from prominent known lines. We also tested the impact of excluding all wavelengths $\lambda>1.33\,\mu$m to avoid bias from the putative water absorption found in two LRDs \citep{Wang2026}, but ultimately, this choice did not affect our conclusions. 
We specifically evaluated the ratio of predicted Paschen and Brackett spectral continua at 0.95\,$\mu$m as this is the shortest wavelength that is directly probed by the Brackett continuum before strong Paschen emission lines obscure the continuum in \sirocco models. For all LRDs, we also fitted blackbody continua to the same wavelengths as for the Paschen and Brackett continua to estimate the blackbody temperature, $T_{\rm BB}$. These blackbody fits provided a mean and spread in reduced chi-squared $\chi_{\nu,\,\mathrm{BB}}^2 = 4.7\pm7.6$ compared to $\chi_{\nu,\,\mathrm{poly}}^2 = 1.9\pm1.6$ for the polynomial parametrization. After inflating each object's uncertainties so that the polynomial fit yielded a reduced chi-square of unity, we ruled out the null hypothesis that blackbody fits accurately describe the data in 5 (2) out of 14 objects with a significance threshold of 2$\sigma$, or a \(p\) value of $4.5$\% (5$\sigma$, or a \(p\) value of $5.7\times10^{-5}$\,\%). Fig.~\ref{fig:paschen_sample} clearly shows that blackbody models are limited in simultaneously fitting the Paschen and Brackett continuum in the lowest- and highest-redshift objects in the sample. The lower S/N spectra result in correspondingly larger uncertainties in the polynomial coefficients and in the extrapolation, and they ultimately provide little discernible statistical power on the nature of the continuum. More generally, we emphasise that physically motivated modifications to blackbody emission can be invoked even for highly thermalised emission \citep[e.g.\ H$^-$-opacity kink, broader and narrower SEDs, see][]{Liu2026}, but the key observational test introduced by this paper is the sudden change in spectral slope at the Paschen limit, which is an observational difficulty for any smoothly varying thermal-like emission. In Fig.~\ref{fig:sirocco_ex} we show a piecewise linear function with a free-to-vary break position, fit simultaneously to both continua. This illustrates that the expected change in spectral slope is near the Paschen limit, analogous to the change in spectral slope near the Balmer limit.

Second, we modelled the continuum as a reddened \pyneb template, $F_{\nu}\propto C^{\rm PyNeb}_{\nu}(T_e, n_e)\times10^{-0.4 A_\lambda}$, where $A_\lambda$ is the extinction curve, and $C^{\rm PyNeb}_{\nu}$ is the numerically evaluated continuum emissivity for a given electron temperature and density \citep{Luridiana2015}. In practice, this template represents the optically thin limit of this nebular recombination emission and not a fully self-consistent radiative-transfer model such as in \sirocco. Under this assumption, and adopting a prior of $n_e\in[10^6{-}10^{10}]\,\mathrm{cm^{-3}}$, we were able to extract the electron temperature and dust reddening from the continuum spectral slope. 
In optically thin conditions, the Paschen jump prominence is a strong complementary diagnostic of the electron temperature as it is a local emissivity diagnostic. However, the jump prominence is obscured in the \sirocco models observationally by the strong emission of blended Paschen lines and intrinsically because in optically thick conditions not only the emission coefficient, $j_\nu$, but also the absorption coefficient, $\kappa_\nu$, changes over the Paschen limit. Thus, in the \sirocco models, while the microscopic opacity/emissivity edge still exists, the emergent discontinuity from the source function $S_\nu=j_\nu/\kappa_\nu$ is weak and perhaps only clearly apparent when the atomic levels are removed that contribute the emissivity just above the Paschen limit (see $n\leq10$ in Fig.~\ref{fig:LRD_example}).

\begin{figure}
\begin{center}
    \includegraphics[angle=0,width=0.5\textwidth,viewport=12 10 380 380, clip=]{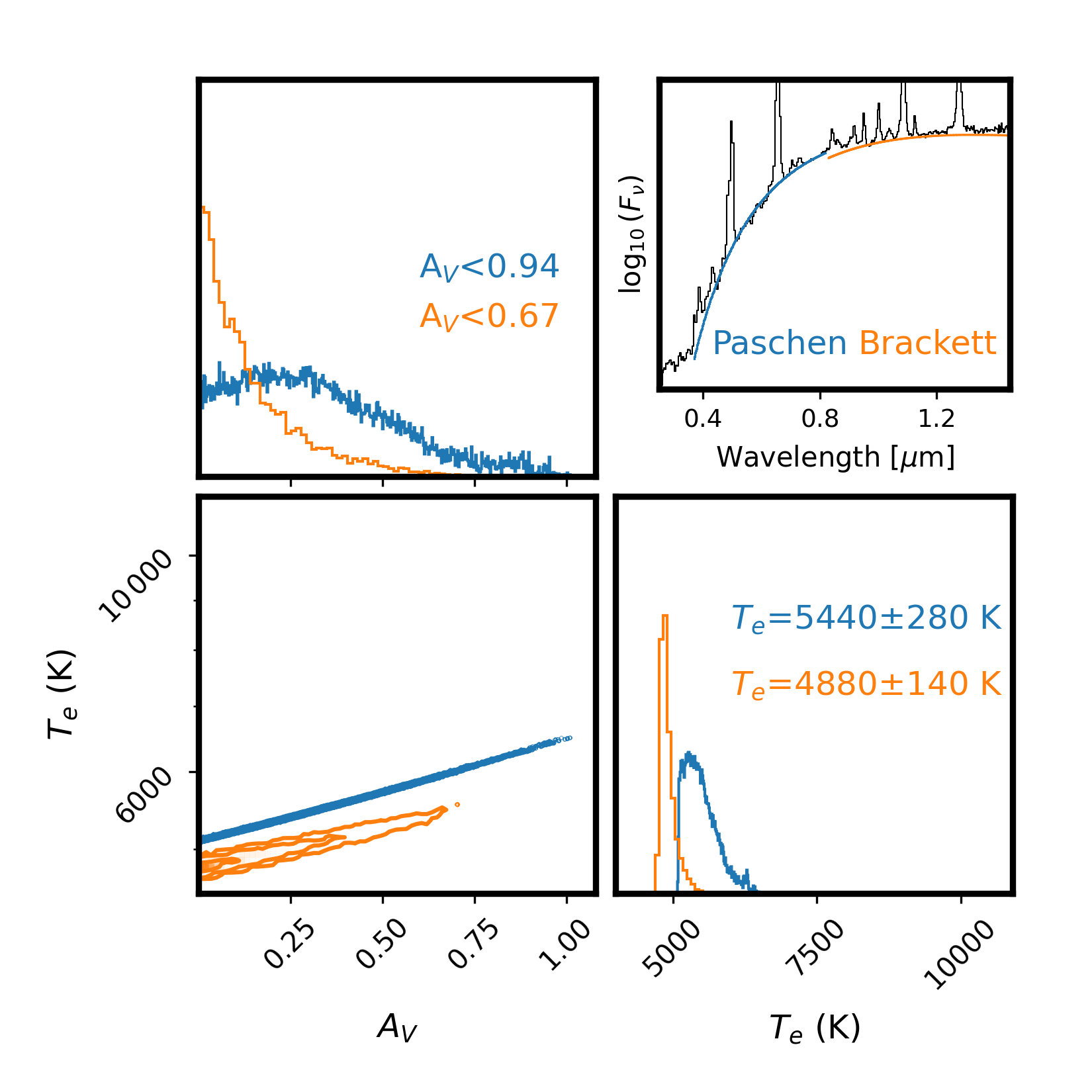}
\end{center}
\vspace{-0.6cm}
\caption{ Constraints on the electron temperature, $T_e$, and dust extinction, $A_V$, for the Rosetta stone from the Paschen and Brackett spectral slope assuming the origin as free-bound emission with \pyneb. The top right panel shows the corresponding $F_\nu(\lambda)$ spectral slope in \pyneb models and the data. The Paschen and Brackett continua favour a modest reddening ($A_V\lesssim0.7\text{-}0.9$ mag, 3$\sigma$ limit) and low-temperature solution, $T_e\sim5000\,\mathrm{K}$, which was similarly required for the e-folding width found in the broad electron-scattered lines \citep{Sneppen2026}. }
\label{fig:electron_temp}
\end{figure}

\begin{figure}
\begin{center}
    \includegraphics[angle=0,width=0.5\textwidth]{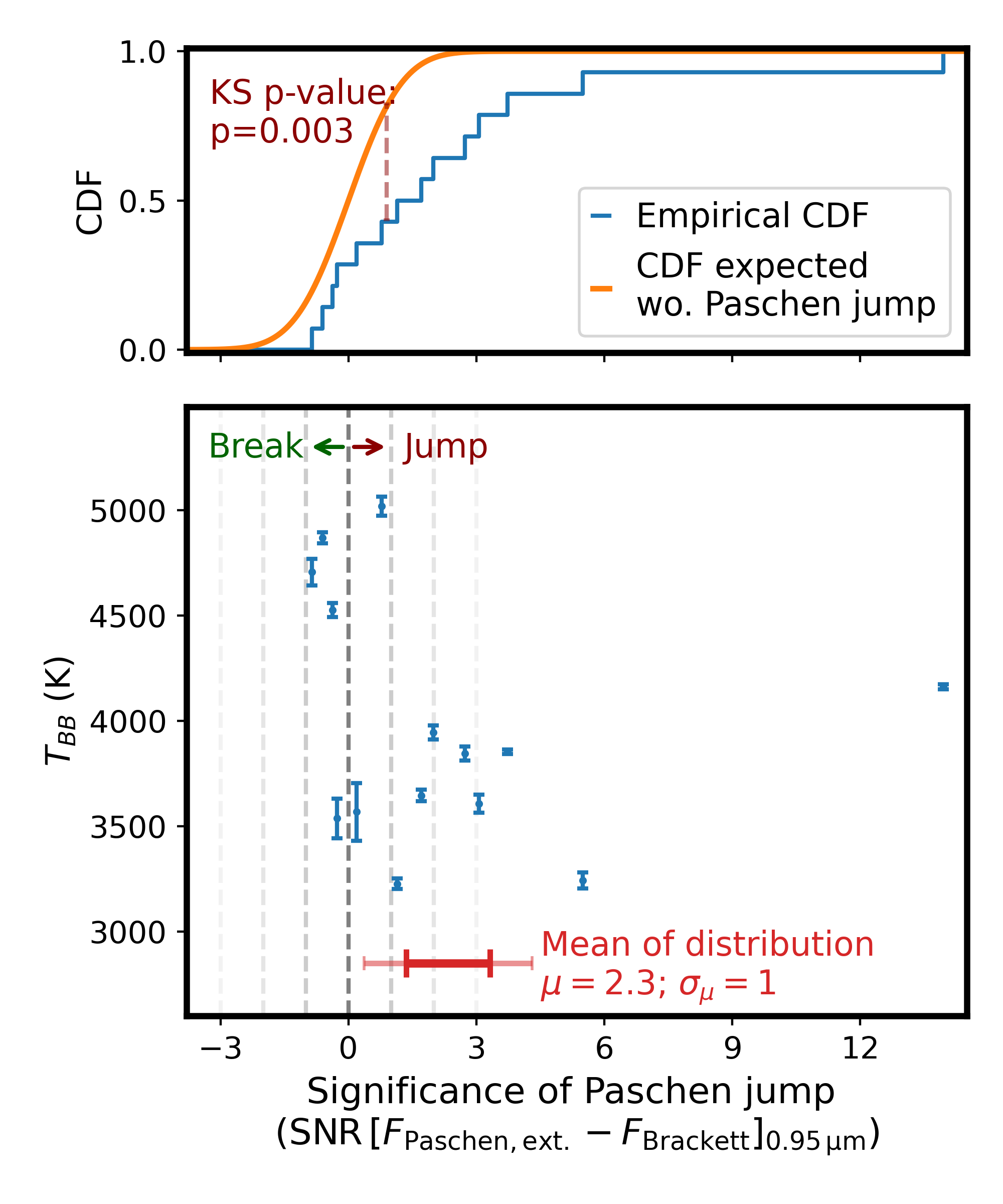}
\end{center}
\vspace{-0.6cm}
\caption{Spectra display systematic evidence of a Paschen jump, quantified from the difference between Paschen- and Brackett-extrapolated continua at 0.95\,$\mu$m. Bottom panel: Associated S/N estimated from posterior samples of the continuum fits for each object (uncertainties from the 16th–84th percentile range at 0.95\,$\mu$m) and plotted against the best-fit blackbody temperature. Four out of 14 objects show Paschen jump signatures at $>3\sigma$. Top panel: Cumulative distribution function of the significance of jumps (blue) compared to the null expectation in the absence of a Paschen jump signature (orange). The Kolmogorov–Smirnov (KS) test indicates that the two distributions are inconsistent with being drawn from the same parent distribution ($p<0.05$). }
\label{fig:significance}
\end{figure}

\begin{figure}
\begin{center}
    \includegraphics[angle=0,width=0.5\textwidth]{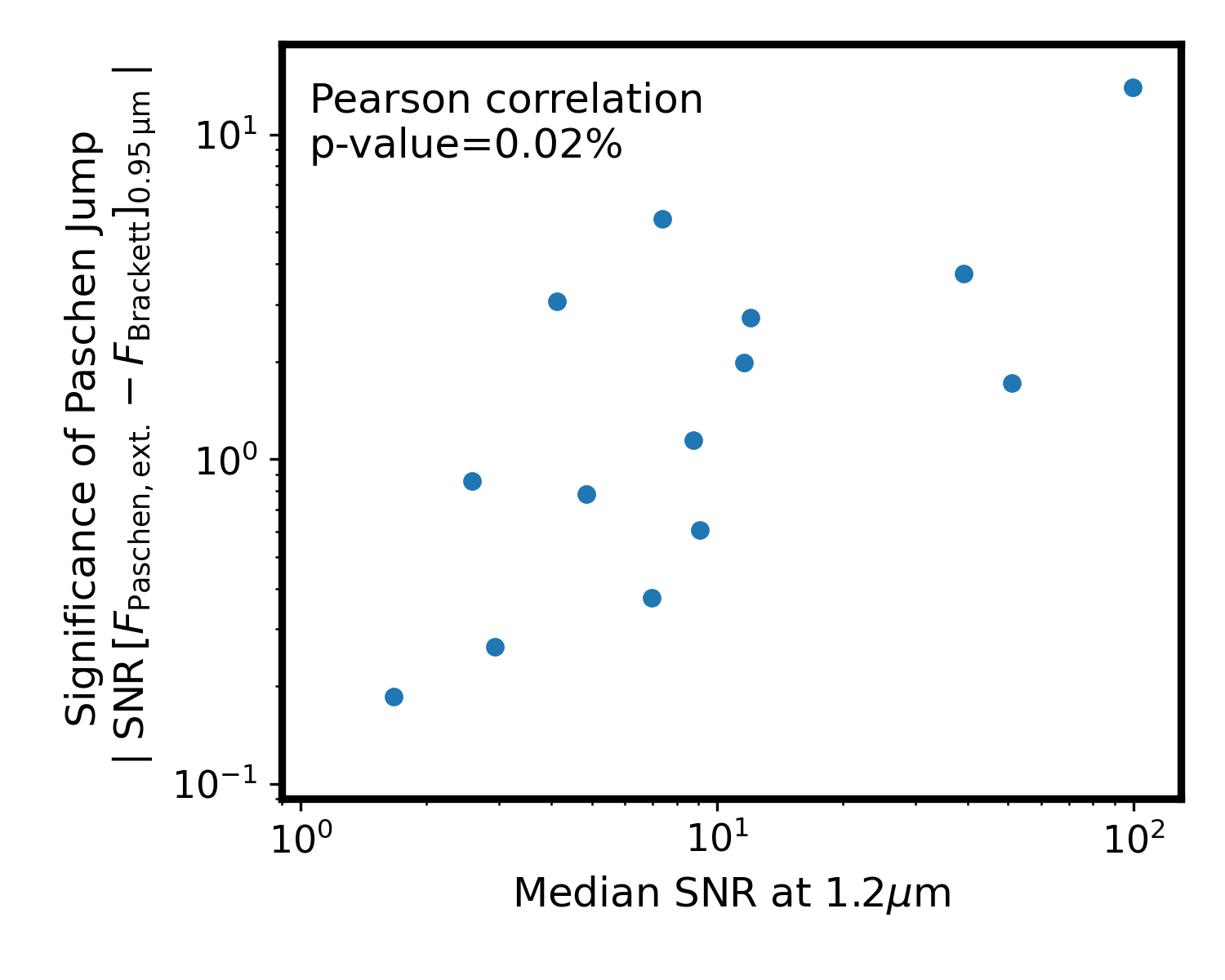}
\end{center}
\vspace{-0.6cm}
\caption{Significance of the Paschen discontinuity as a function of the median S/N in the Brackett continuum, measured over \(\lambda/\mu{\rm m}\in[1.15,1.25]\). The Pearson test gives \(p=2\times10^{-4}\) for the correlation between |Paschen jump significance| and S/N, indicating that the detectability of the jump is strongly S/N-dependent. This implies that low-S/N spectra have limited diagnostic power for identifying or constraining the Paschen discontinuity.}
\label{fig:SNR_SNR}
\end{figure}

\begin{figure*}
\sidecaption
\includegraphics[width=12cm,viewport=21 10 585 480, clip=]{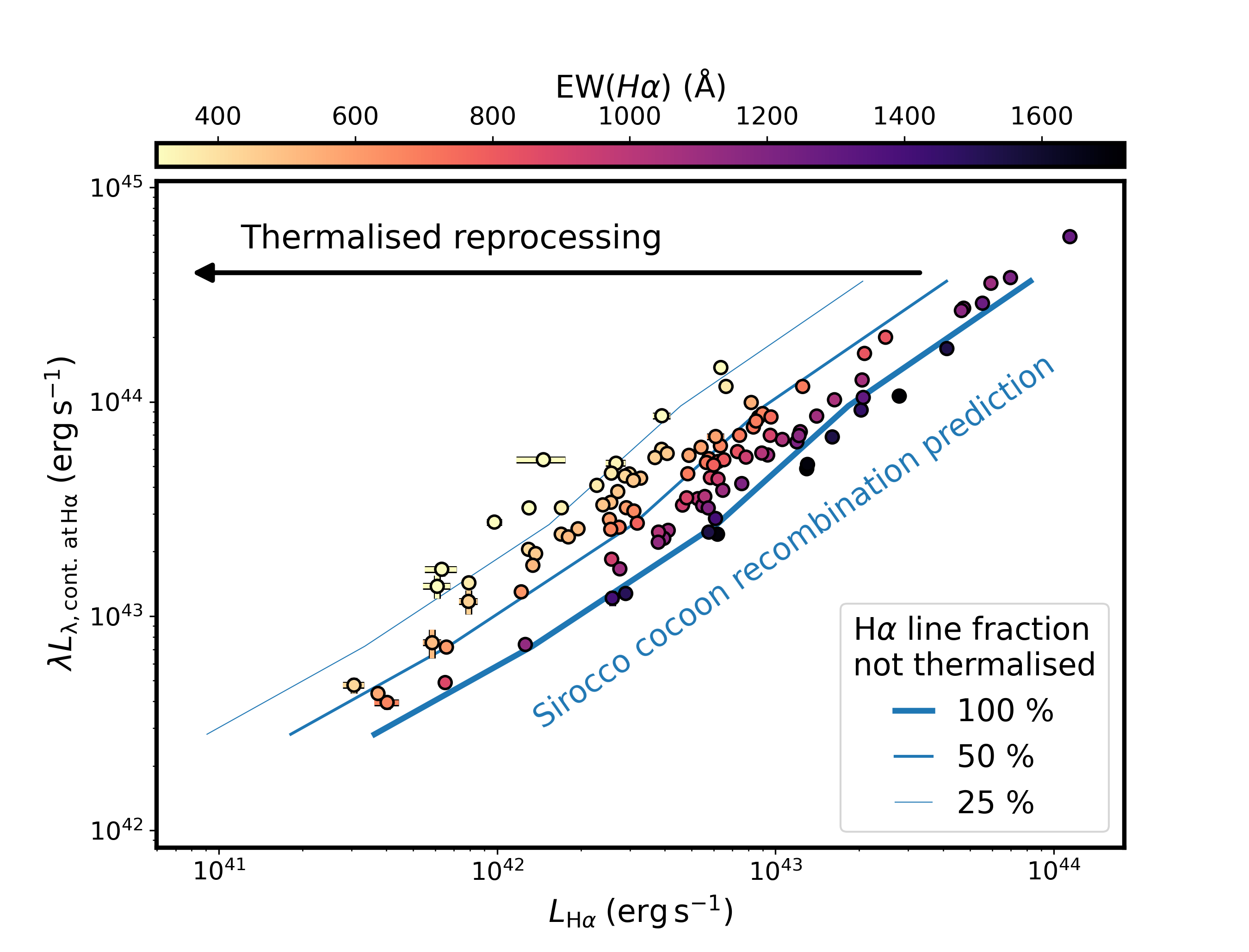}
\caption{Comparison of the H$\alpha$ line luminosity (integrated above the continuum) and the continuum luminosity at the H$\alpha$ wavelength for the LRD sample from \cite{deGraaff2025b}. In a \sirocco\ model sequence from \cite{Sneppen2026} (thick blue curve), these quantities are tightly linked because the continuum and the line emission arise from recombination processes (free-bound recombination and recombination cascades, respectively). Any additional blackbody-like reprocessing moves objects away from this relation by reducing the emergent line luminosity relative to the optical continuum; the offset increases with the assumed fraction of reprocessing (thinner blue curves). In the limit of a optically thick envelope and a thermalised blackbody outside the line-forming region, the continuum-subtracted H$\alpha$ line luminosity, $L_{\mathrm{H}\alpha}$, tends to zero, because the thermalised continuum is not itself hot enough for the requisite photoionisation.}
\label{fig:blackbody_sim}
\end{figure*}


\section{Results}\label{sec:results}
Four of 14 objects significantly deviate from a smooth rollover at a significance of $>3\sigma$, with the Paschen continuum extrapolation lying consistently above the Brackett continuum (see legend in Fig.~\ref{fig:paschen_sample} or Fig.~\ref{fig:significance}). Notably, this includes a $\sim14\sigma$ outlier in a relatively modest sample size of objects. 
This is the Rosetta Stone LRD, which has the highest continuum S/N in the sample with $\mathrm{S/N_{1.2\,\mu m}}\approx100$. We therefore highlight this in the top panel of Fig.~\ref{fig:paschen_sample}, where we show for comparison multiple blackbodies of varying temperature and subject to differing levels of dust reddening. Although simple blackbody emission can reproduce the Brackett continuum or the very bluest Paschen continuum emission, they cannot smoothly fit both wavelength regimes. 

Most objects consistent with smooth rollovers come from the lower-S/N part, while none of the five objects with $\mathrm{S/N_{1.2\,\mu m}}>10$ (useful to constrain the Brackett continuum) is consistent with smooth rollover within 1$\sigma$, and in all cases because the Paschen extrapolation overshoots the Brackett continuum. 
Conversely, the low-S/N part of the sample does not display significant evidence of Paschen jumps, although it also does not rule out their presence. The significance of the Paschen jump and the S/N evaluated at 1.2$\mu$m, that is, in the Brackett continuum used to constrain the redward continuum, are highly correlated (see Fig.~\ref{fig:SNR_SNR}): the Pearson p value for the null hypothesis of no correlation is $p=2\times10^{-4}$. Since all of the higher-S/N objects are significantly discrepant from a smooth continuum, we conclude that nebular emission is likely common within the LRD population, but not necessarily universal. However, in Sect.~\ref{sec:EW}, we show that another signature of nebular emission, extremely high equivalent-width line emission, is universal for LRDs.

While a Paschen jump signature is inconsistent with the null hypothesis of a cold thermalised blackbody continuum, we emphasise that an apparently smooth rollover (i.e.\ the absence of a clear Paschen jump) is consistent with direct recombination emission in the partially optically thick regime seen in \sirocco models. However, in the plot of the significance of a Paschen spectral break versus the best-fit blackbody temperature in Fig.~\ref{fig:significance}, four $T_{\mathrm{BB}}\gtrsim5000\,\mathrm{K}$ LRDs are consistent with a smooth rollover. These resemble `the Cliff'-like LRDs \citep{deGraaff2025}. If interpreted literally as blackbody temperatures, these $T_{\rm BB}$ values fall in a regime in which the H$^-$-opacity can be efficient in stellar atmospheres, potentially producing pseudo-thermalised continua. Reproducing blackbody-adjacent continuum shapes may motivate the inclusion of H$^-$-opacity within \sirocco in the future.

\subsection{Spectral slope constraints on $T_e$}
Using the nebular code \pyneb to fit the electron-temperature, $T_e$, we found that the detailed shape of the Paschen continuum and the overall rise towards the Paschen series limit are consistent with $T_e\lesssim10^4\,K$ and low reddening, $A_V\lesssim1$, across the sample. For instance, for the Rosetta stone, the Paschen and Brackett continuum slopes would independently suggest $5400\pm300$ K and $4900\pm100$ K, respectively, for a flat prior on the dust extinction up to $A_V=5$ (see Fig.~\ref{fig:electron_temp}). The dust extinction from this estimate is $A_V\lesssim0.7\text{-}0.9$ mag as a $3\sigma$ upper limit, which is modest in comparison to stellar UV fits and narrow Balmer line ratios assuming case B recombination \citep[respectively, $A_V\sim0.68$ and $A_V=1.1\pm0.1$,][]{Juodzbalis2024_rosetta}, although we note that other opacity sources or emission components might bias the spectral shape.
Generally, the inferred $T_e$ and $A_V$ in the sample are strongly degenerate as they similarly steepen the spectral slope \citep{Guseva2006}. The minimum Balmer decrement is that of case B recombination ratios, and based on this, we can invoke a prior on $A_V$ with an upper limit ranging from $A_V\sim1$ to $A_V\sim5.2$ depending on the object in the sample. Invoking this prior means that all five well-constrained objects are only consistent with $T_e\leq10^{4}$\,K. While the remaining objects permit a broader family of reddening and temperatures, none are inconsistent with $T_e\leq 10^4$\,K. 

In reality, there is likely to be some stratification in temperature within the cocoon, as is indeed found in the Sirocco models. However, the modest gas temperatures from our single-temperature model are consistent with those implied by Fe\,\textsc{ii} emission \citep[$\sim\!7000$\,K,][]{Torralba2025}, Ca triplet absorption \citep[$\sim\!5000$\,K,][]{Lin2025}, and electron-scattering widths \citep[$\sim\!5000$\,K,][]{Sneppen2026}. This consistency with the electron-scattering widths indicates that the recombining gas that produces the continuum and the gas that causes the line broadening share similar thermodynamic conditions. In turn, this strengthens the argument that the line-forming region is approximately co-spatial with the continuum-emitting region. 

\subsection{A test of thermalised reprocessing: ${\rm EW(H\alpha)}$}\label{sec:EW}

In the LRD population, continuum luminosity and hydrogen line emission are tightly related \citep{deGraaff2025b,Pang2026,Yanagisawa2026b}, as predicted by our radiative transfer models, because the continuum and lines both arise from recombination emission in the same general gas region (see Fig.~\ref{fig:blackbody_sim}). From this perspective, we also caution against interpreting the emission as a true blackbody. Strong hydrogen recombination lines are produced in the inner ionised cocoon, whereas the continuum might subsequently undergo additional reprocessing in cooler partially ionised outer layers mediated by H$^-$-opacity. Because the continuum is already smooth and thermal-like, additional absorption and re-emission in optically thick outer layers would mostly preserve its broad spectral character. By contrast, line emission is initially concentrated into spectral features, so repeated reprocessing would tend to convert part of that line luminosity into a smoother continuum-like component. If the outer layers became sufficiently optically thick to thermalise the radiation field, the emergent spectrum seen by an external observer would therefore show diluted line features, even though the continuum luminosity is similar. In that limit, the observable coupling between line strength and continuum luminosity is reduced in proportion to the assumed level of reprocessing/thermalisation. 

The line-to-continuum ratio, $\mathrm{EW}(H\alpha)={L_{H\alpha}}/{L_{\rm \lambda,cont.\,at\,H\alpha}}$, is high, 
reaching equivalent widths of $\mathrm{EW(H\alpha)_{LRD}}=1700\pm100$\,\AA, which is well above the expectation from standard AGN relations \citep[i.e.\ $\mathrm{EW(H\alpha)_{AGN}\sim200\,\AA,}$][]{Greene2005,Maiolino_Chandra_2024}. 
The prominent jump in an object such as CAPERS 6368-30440 is consistent with direct line emission with EW(H$\alpha)=1500\pm60$\,\AA. The subtler jump in the Rosetta stone with EW(H$\alpha)=820\pm6$\,\AA\ might indicate modest reprocessing to the continuum, while the inconclusive evidence of a jump in Uncover 2561-20698 is matched by the weakest line-emission of the sample with EW(H$\alpha)=370\pm10$\,\AA. This might motivate a population trend that the nebular signatures of continuum (i.e., Paschen jumps) and lines (i.e. high EW) are related because co-production or high columns reprocess both signatures. However, the limited sample size used here is not sufficient to robustly establish such a correlation, where the S/N dominates the Paschen jump detectability. A high EW alone does not uniquely require a nebular continuum and could also be explained with direct disc emission and a covering factor near unity \citep{Maiolino_Chandra_2024,Yanagisawa2026b}. Regardless, in the Compton-thick limit, this would reduce to the models presented here. 

The thick blue line in Fig.~\ref{fig:blackbody_sim} shows this luminosity relation in the \sirocco model sequence from \cite{Sneppen2026}, in which the density is varied up to a Balmer-break strength of $D_{4000}=F_{\lambda=4200\,Å}/F_{\lambda=3500\,Å}=20$. Any additional thermalisation in exterior layers would not dramatically change the continuum colours and would mainly act to weaken the apparent line emission relative to the continuum by a factor of a few. We indicate the approximate scale of this effect with thinner blue lines, which are simple offsets from the main \sirocco relation and not separately computed models.

Thus, i) the tight observed correlation and ii) the high equivalent width found in LRDs both suggest that thermalisation processes in any putative stellar atmosphere are limited and do not fully reprocess the incident emission. Even the strongest outliers from the fiducial \sirocco recombination emission relation are a factor of a few underluminous in the \ha line at most, such as objects like `the Cliff'. 
Alternatively, broad lines superposed on a thermalised continuum might be produced if the broad permitted lines were formed outside the black hole envelope, but this would require an ionising energy source distinct from the AGN that is producing the optical-IR continuum \citep[e.g.][]{Inayoshi2025b} or would necessitate an inverted temperature structure with a highly ionised exterior. Furthermore, given the strong correlation between lines and continuum, this scenario would require the energetic source outside the envelope to be closely tuned to the SMBH accretion luminosity. 
In addition, the observed \ha EW is already very large, so invoking a substantial continuum contribution from any component other than the line-emitting region would require an even more extreme intrinsic EW in that region. 
Lastly, given that the electron scattering and absorption is acting on broad H and He lines, it would also still require this external broad-line component to lie within some extended high-column density envelope, which also has a $T_e$ very similar to that of the continuum.

\subsection{A single nebular component producing lines and continua}
A key success of the \sirocco LRD spectra is thus their ability to reproduce the continuum, line strengths, and line profiles simultaneously, without requiring multiple separately fitted components. The model naturally produces a thermal-looking optical continuum, with $k_BT_{\rm BB}\!\sim\!0.3\,{\rm eV}$, while also predicting recombination lines from species requiring much higher ionisation energies deeper within the cocoon, such as \Lya, \ha, and \hei\,$\lambda 1083.3$\,nm. Moreover, the column density implied by the Balmer break agrees with that inferred from electron-scattering broadening \citep{Sneppen2026}. Likewise, the electron temperature inferred from the continuum shape agrees without fine-tuning with that inferred from scattering-broadened line widths. 

Notably, nebular emission also predicts a corresponding population of Balmer jump objects, which are expected to be similar to LRDs, but without the strong Balmer absorption signatures \citep{Sneppen2026}. These objects are predicted to have blue continuum colours with a Balmer jump, exponential broad wings due to electron scattering, and X-ray weakness due to Compton-thick gas \citep{Sneppen2026d}. They thus share similar observational qualities with the so-called LBD population \citep{Brazzini2026}. 

\begin{acknowledgements}
The \sirocco code \citep{Matthews2025} and documentation are publicly available\footnote{\href{https://github.com/sirocco-rt/sirocco}{https://github.com/sirocco-rt/sirocco}}. The comparison to observed LRD makes use of the public \jwst data collected as part of several observational programs with the NIRSpec spectrograph \citep{Jakobsen2022} as enumerated in the original sample paper by \cite{deGraaff2025b} with the following PIDs; \citep[CANUCS][]{Sarrouh2026}; 6368 (CAPERS); 1345 \citep[CEERS][]{Finkelstein2025}; 2750 (PI: Arrabal Haro); 2767 (PI: Kelly); 6585 (PI: Coulter); 1433 (PI: Coe); 2198 (PI: Barrufet); 4106 (PIs: Nelson and Labbe); 1211, 1212, 1213, 1215 \citep[GTO-WIDE][]{Maseda2024}; 1180, 1181, 1286 \citep[JADES][]{eisenstein2023a,Bunker2023,Deugenio2024,ArrabalHaro2021}; 4233 \citep[RUBIES][]{DeGraaff2024_RUBIES}; 5224 \citep[MoM][]{Naidu2025}; 5105 \citep[5105][]{Shen2024}; 2561 \citep[2561][]{Bezanson2024,Price2025}. These observations have been uniformly reduced and published as part of the Dawn \jwst Archive\footnote{\href{https://dawn-cph.github.io/dja}{https://dawn-cph.github.io/dja}} (DJA), \cite{DeGraaff2024_RUBIES,Heintz2025,Pollock2025}. DJA is an initiative of the Cosmic Dawn Center (DAWN), which is funded by the Danish National Research Foundation under grant DNRF140. 
AS, DW and KEH acknowledge funding by the European Union (ERC, HEAVYMETAL, 101071865). Views and opinions expressed are, however, those of the authors only and do not necessarily reflect those of the European Union or the European Research Council. Neither the European Union nor the granting authority can be held responsible for them. AS and KEH acknowledge support from Independent Research Fund Denmark (DFF) under grant 6219-00033B and 5251-00009B, respectively. JHM acknowledges funding from a Royal Society University Research Fellowship (URFR1221062). SAS were supported by the UK’s Science and Technology Facilities Council (STFC, respectively grant ST/V001000/1 and ST/X00094X/1). 
\end{acknowledgements}

\bibliographystyle{aa}
\bibliography{refs} 

\setcounter{section}{1}
\setcounter{equation}{0}
\setcounter{figure}{0}
\renewcommand{\thesection}{Appendix \arabic{section}}
\renewcommand{\theequation}{A.\arabic{equation}}
\renewcommand{\thefigure}{A.\arabic{figure}}

\end{document}